\documentclass[12pt]{article}
\usepackage{amssymb,amsfonts}
\def \la{\lambda}

\def \sh{{\rm sinh}}

\def \t{\tau}

\newcounter{map}
\newcounter{fel}
\newcounter{aw}
\newcounter{aww}
\begin{document}
\baselineskip 18pt

\title{$R$-matrices for integrable $SU(2)\times U(1)$-symmetric $S=\frac{1}{2}$ spin-orbital
chains}
\author{P.~N.~Bibikov\\ \it V.~A.~Fock Institute of Physics,
Sankt-Petersburg State University}

\maketitle

\vskip5mm

\begin{abstract}
The Yang-Baxter equation for a $SU(2)\times U(1)$-symmetric $S=\frac{1}{2}$ spin-orbital chain
was solved using the special computer algorithm developed by the author.
The 7 new $R$-matrices separated on 4 groups are presented. Among the obtained integrable models
there are special cases related to 1D ferromagnet ${\rm TDAE-C}_{60}$, 1D
superconductors ${\rm AC}_{60}$ (A=K, Cs, Rb), the quarter filled ladder compound
${\rm NaV}_2{\rm O}_5$ and the model of correlated electrons on a chain of Berry phase molecules.
\end{abstract}

\begin{section}
{Introduction}
\end{section}
At the beginning of 1970-s Kugel and Khomskii \cite{1} and independently S. Inagaki \cite{2}
suggested two various models to describe magnetic properties of solids with orbital degeneracy in
electron systems of atoms. Starting from the two-band Hubbard model \cite{3} they have obtained
low-energy Hamiltonians depending on both spin and pseudospin (orbital) operators.
Kugel and Khomskii took into account geometry
of d-orbitals entailing to asymmetry of hopping integrals and obtained a general but realistic
Hamiltonian. On the contrary S. Inagaki postulating the symmetric hopping
have distinguished between the Coulomb repulsions on the same and different orbitals.
It was suggested in \cite{1}-\cite{3} that a
non-trivial coupling between spin and orbital terms would result to a complex magnetic behavior
such as a ferromagnetism induced by orbital ordering.

While the spin dependence of the Hamiltonians \cite{1},\cite{2} has purely $SU(2)$-invariant
Heisenberg form its dependence upon pseudospin is more complicated and
in the simplest case possess only the $U(1)$ symmetry related to
rotations along the $z$ axis in the pseudospin space.

After the works \cite{1},\cite{2} some new applications of $SU(2)\times U(1)$-invariant spin models
were suggested in a number of papers \cite{4}-\cite{10}. Ground state and excitations of some
$SU(2)\times U(1)$-invariant spin chains were studied in \cite{11},\cite{12}.

Suggested by L. D. Faddeev and his school Quantum Inverse Scattering Method
(Algebraic Bethe Ansatze) \cite{13}-\cite{15} is the most elaborated approach for exact detailed
analysis of a one-dimensional integrable spin chain.
The latter is described by a Hamiltonian,
\begin{equation}
\hat H=\sum_{n=1}^NH_{n,n+1},
\end{equation}
acting on the finite-dimensional space $({\mathbb C}^M)^{\otimes N}$ ($M=2,3,4,..$).
The each term $H_{n,n+1}$ acts nontrivially as a $M^2\times M^2$ matrix $H$
only on the tensor product of two neighbor spaces ${\mathbb C}^M_n\otimes{\mathbb C}^M_{n+1}$.

The keystone of this approaches is the Yang-Baxter equation:
\begin{equation}
R_{12}(\la-\mu)R_{23}(\la)R_{12}(\mu)=R_{23}(\mu)R_{12}(\la)R_{23}(\la-\mu),
\end{equation}
with the initial regularity condition:
\begin{equation}
R(0)=cI.
\end{equation}
Here $R(\lambda)$ is a $M^2\times M^2$ matrix, $I$ is the matrix unity while $c$ is an arbitrary
nonzero constant.

If the Hamiltonian density matrix $H$ relates to $R(\lambda)$ by the following formula:
\begin{equation}
H=\frac{\partial R(\la)}{\partial\la}|_{\la=0},
\end{equation}
then the system (1) is integrable.

In \cite{16} a new method was suggested for solving the Eqs (2),(3). It is based on the series
expansion for $R$-matrix:
\begin{equation}
R(\la)=\sum_{n=0}^{\infty}\frac{1}{n!}R^{(n)}\la^n,
\end{equation}
where,
\begin{eqnarray}
R^{(1)}&=&H,\nonumber\\
R^{(2)}&=&H^2,\nonumber\\
R^{(3)}&=&H^3+K,\nonumber\\
R^{(4)}&=&H^4+2(HK+KH),\nonumber\\
R^{(5)}&=&H^5+L+2(KH^2+H^2K)+6HKH,\nonumber\\
R^{(6)}&=&H^6+KH^3+H^3K+9(H^2KH+HKH^2)+\nonumber\\
&&\qquad\qquad\qquad\qquad\qquad+10K^2+3(HL+LH).
\end{eqnarray}
The matrices $K$ and $L$ may be obtained from the following integrability conditions
\cite{13},\cite{16},\cite{17}:
\begin{eqnarray}
K_{23}-K_{12}&=&{[}H_{12}+H_{23},{[}H_{12},H_{23}{]]},\nonumber\\
L_{23}-L_{12}&=&[H_{12}^3+H_{23}^3+3(K_{12}+K_{23}),J]+
3(H_{12}[J,H_{12}]H_{12}+
H_{23}[J,H_{23}]H_{23})+\nonumber\\
&&(H_{12}H_{23}+H_{23}H_{12})(K_{23}-K_{12})+(K_{23}-K_{12})(H_{12}H_{23}+
H_{23}H_{12})-\nonumber\\
&&2(H_{12}(K_{23}-K_{12})H_{23}+H_{23}(K_{23}-K_{12})H_{12}),
\end{eqnarray}
where $J=[H_{12},H_{23}]$.

As the Eq.(2) as the Eqs (6) and (7) are invariant under the following $Q\in SL(M)$ action:
\begin{equation}
X\rightarrow Q^{-1}\otimes Q^{-1}XQ\otimes Q,
\end{equation}
($Q\in SL(M)$, $X=H,K,R(\lambda)$.

It was shown in \cite{16} a detailed analysis of series expansions applied to quotients of
the $R$-matrix entries gives possibility to guess right the whole $R$-matrix. An alternative
approach for solving the system (2),(3) is given in the recent paper \cite{19}.

In the next sections
we shall present the obtained by our approach $R$-matrices related to
the general $SU(2)\times U(1)$-symmetric spin-orbit Hamiltonian,
\begin{eqnarray}
H_{n,n+1}&=&(s_ns_{n+1})(a_1+a_2(\t^x_n\t^x_{n+1}+\t^y_n\t^y_{n+1})+a_3\t^z_n\t^z_{n+1}+
\frac{1}{2}a_6(\tau_n^z+\tau_{n+1}^z))+\nonumber\\
&+&a_4(\t_n^x\t_{n+1}^x+\t_n^y\t_{n+1}^y)
+a_5\t^z_n\t^z_{n+1}+\frac{1}{2}a_7(\tau_n^z+\tau_{n+1}^z),
\end{eqnarray}
which may be parameterized by the set of coefficients: $S=\{a_1,a_2,...,a_7\}$. Here in (9) the spin
and pseudospin operators are expressed from the Pauli matrices:
\begin{equation}
s^k=\frac{1}{2}\sigma^k\otimes I_2,\quad \tau^k=\frac{1}{2}I_2\otimes\sigma^k,\quad k=x,y,z.
\end{equation}

The following change of coefficients:
\begin{equation}
\{a_1,a_2,a_3,a_4,a_5,a_6,a_7\}\rightarrow\{a_1,a_2,a_3,a_4,a_5,-a_6,-a_7\},
\end{equation}
does not destroy an integrability or change the spectrum of Hamiltonian because it
corresponds to the transformation (8) with $Q=I_2\otimes\sigma^x$. .

For a chain with even numbers of sites the same is true for the following change of variables:
\begin{equation}
\{a_1,a_2,a_3,a_4,a_5,a_6,a_7\}\rightarrow\{a_1,-a_2,a_3,-a_4,a_5,a_6,a_7\},
\end{equation}
which corresponds to the graduated version of (8).
\begin{equation}
H_{2n,2n+1}\rightarrow4\tau_{2n}^zH_{2n,2n+1}\tau_{2n}^z,\quad
H_{2n-1,2n}\rightarrow4\tau_{2n}^zH_{2n-1,2n}\tau_{2n}^z.
\end{equation}

The case,
\begin{equation}
S^{KH}=\{1-\alpha,0,4(1+\alpha),0,1+\alpha,4,-1\}
\end{equation}
corresponds to the Kugel-Khomskii model of 1D perovskite \cite{1}.

The case,
\begin{equation}
S^{I}=\{\frac{1}{2}(\alpha-\beta)+\gamma,2(\alpha+\beta),2(2\gamma+\beta-\alpha),
\frac{1}{2}(3\beta-\alpha),\frac{1}{2}(3\beta+\alpha-2\gamma),0,0\},
\end{equation}
(where unphysical region $\alpha\approx\beta\approx\gamma$, but $\beta>\alpha$ and $\beta>\gamma$)
corresponds to the Inagaki's model \cite{2}.
In \cite{4} it was also applied to organic 1D ferromagnet ${\rm TDAE-C}_{60}$ and in \cite{5}
to the family of 1D superconductors $AC_{60}$ ($A={\rm K,Cs,Rb}$) (with $T_c>30K$)

The special case of (15) with,
\begin{equation}
\alpha =1-\delta,\quad\beta=1+\delta,\quad\gamma=1-\delta^2,
\end{equation}
(where in physical region $0<\delta<1$) was studied in \cite{6}.

The two cases,
\begin{eqnarray}
S_1^{{\rm NaV}_2{\rm O}_5}&=&\{\alpha,8\beta,4\alpha,2\beta,4\beta-\alpha,0,0\},\\
S_2^{{\rm NaV}_2{\rm O}_5}&=&\{-\alpha,4\alpha,4\alpha,3\alpha,3\alpha,-4\beta,\beta\},
\end{eqnarray}
correspond to limiting cases of the model describing the quarter-filled ladder
compound $\alpha'-{\rm NaV}_2{\rm O}_5$ \cite{7}.

The case:
\begin{equation}
S^{st}=\{1,2,0,0,0,0,0\},
\end{equation}
corresponds the effective spin-tube Hamiltonian suggested in \cite{8}.

Let us also mention the paper \cite{9} where the spin-orbital Hamiltonian was applied to
arrays of quantum dots.

In the special $SU(4)$-symmetric point:
\begin{equation}
S^{SU(4)}=\{1,4,4,1,1,0,0\},
\end{equation}
the model (20) was solved in \cite{18} ($R(\la)=\eta I+\lambda{\cal P}$).
This point corresponds
to degenerative cases of (15) ($\alpha=\beta=\gamma$) (or $\delta=0$)
Using the transformation (12),(13) we may obtain the $R$-matrix for the model with
\begin{equation}
\tilde S^{SU(4)}=\{1,-4,4,-1,1,0,0\}.
\end{equation}

In the special $Sp(4)$-symmetric point:
\begin{equation}
S^{Sp(4)}=\{1,4,8,2,1,0,0\},
\end{equation}
(equivalent to $\tilde S^{Sp(4)}=\{1,-4,8,-2,1,0,0\}$) related to $3\alpha=\beta=\gamma$ in (17)
the $R$-matrix was presented in \cite{20}.

Except (20) and (22) no integrable cases of the Hamiltonian (9) were studied up to now.
In order to start this process we have
solved the system (2),(3) related to (9). The calculations were performed by two steps.
On the first using the ${\rm Gr{\ddot o}bner}$ package of the computer algebra
system MAPLE 7 we have found 7 new solutions of the system (7).
On the second we derived the corresponding $R$-matrices using the approach suggested in \cite{16},
\cite{19}.

For convenience of representation the obtained $R$-matrices are separated on 4 groups.
In the each one all $R$-matrices have similar positions of non-zero entries therefore they may
be presented in a unique form.

The author is grateful to P.~P.~Kulish for statement of the problem and to M.~J.~Martins for
helpful discussion.

Everywhere below $\varepsilon=\pm1$.

\section{The group 1}
In this group the $R$-matrix corresponds to
\begin{equation}
S^{(1)}=\{0,4,0,1,0,4\varepsilon,\varepsilon\},
\end{equation}
and has the following form,
\begin{equation}
R^{(1)}(\la)=\left(\begin{array}{cccccccccccccccc}
f_+&0&0&0&0&0&0&0&0&0&0&0&0&0&0&0\\
0&\eta&0&0&g_1&0&0&0&0&0&0&0&0&0&0&0\\
0&0&\eta&0&0&0&0&0&g_2&0&0&0&0&0&0&0\\
0&0&0&\eta&0&0&0&0&0&0&0&0&g_1&0&0&0\\
0&g_1&0&0&\eta&0&0&0&0&0&0&0&0&0&0&0\\
0&0&0&0&0&f_-&0&0&0&0&0&0&0&0&0&0\\
0&0&0&0&0&0&\eta&0&0&g_1&0&0&0&0&0&0\\
0&0&0&0&0&0&0&\eta&0&0&0&0&0&g_3&0&0\\
0&0&g_2&0&0&0&0&0&\eta&0&0&0&0&0&0&0\\
0&0&0&0&0&0&g_1&0&0&\eta&0&0&0&0&0&0\\
0&0&0&0&0&0&0&0&0&0&f_+&0&0&0&0&0\\
0&0&0&0&0&0&0&0&0&0&0&\eta&0&0&g_1&0\\
0&0&0&g_1&0&0&0&0&0&0&0&0&\eta&0&0&0\\
0&0&0&0&0&0&0&g_3&0&0&0&0&0&\eta&0&0\\
0&0&0&0&0&0&0&0&0&0&0&g_1&0&0&\eta&0\\
0&0&0&0&0&0&0&0&0&0&0&0&0&0&0&f_-
\end{array}\right),
\end{equation}
where $f_{\pm}=\pm\varepsilon\lambda+\eta$, $g_1=\lambda$, $g_2=-g_3=\varepsilon\lambda$.

\section{The group 2}
This group consists of two related subgroups. For the first one the $R$-matrix
is the following,
\begin{equation}
R^{(2a)}(\la)=\left(\begin{array}{cccccccccccccccc}
f_+&0&0&0&0&0&0&0&0&0&0&0&0&0&0&0\\
0&\eta&0&0&g_1&0&0&0&0&0&0&0&0&0&0&0\\
0&0&\eta&0&0&0&0&0&g_2&0&0&0&0&0&0&0\\
0&0&0&\eta&0&0&0&0&0&0&0&0&g_1&0&0&0\\
0&g_1&0&0&\eta&0&0&0&0&0&0&0&0&0&0&0\\
0&0&0&0&0&f_-&0&0&0&0&0&0&0&0&0&0\\
0&0&0&0&0&0&\eta&0&0&g_1&0&0&0&0&0&0\\
0&0&0&0&0&0&0&f_-&0&0&0&0&0&0&0&0\\
0&0&g_2&0&0&0&0&0&\eta&0&0&0&0&0&0&0\\
0&0&0&0&0&0&g_1&0&0&\eta&0&0&0&0&0&0\\
0&0&0&0&0&0&0&0&0&0&f_+&0&0&0&0&0\\
0&0&0&0&0&0&0&0&0&0&0&\eta&0&0&g_1&0\\
0&0&0&g_1&0&0&0&0&0&0&0&0&\eta&0&0&0\\
0&0&0&0&0&0&0&0&0&0&0&0&0&f_-&0&0\\
0&0&0&0&0&0&0&0&0&0&0&g_1&0&0&\eta&0\\
0&0&0&0&0&0&0&0&0&0&0&0&0&0&0&f_-
\end{array}\right),
\end{equation}
and there are two solutions. The first one,
\begin{equation}
S^{(2a,1)}=\{1,8\varepsilon,4,2\varepsilon,3,4,-1\},
\end{equation}
corresponds to $f_+=f_-=\lambda+\eta$, $g_1=-g_2=\varepsilon\lambda$. The second,
\begin{equation}
S^{(2a,2)}=\{1,8\varepsilon,4,2\varepsilon,-1,4,3\},
\end{equation}
corresponds to $f_{\pm}=\eta\pm\lambda$, $\varepsilon g_1=g_2=\lambda$.

For the second subgroup the corresponding Hamiltonians may be obtained by from (26),(27)
by the transformation (11) while the $R$-matrices by transposition with respect to the second
diagonal.

\section{The group 3}
In this group $R$-matrices have the form,
\begin{equation}
R^{(3)}(\la)=\left(\begin{array}{cccccccccccccccc}
f_+&0&0&0&0&0&0&0&0&0&0&0&0&0&0&0\\
0&h&0&0&g&0&0&0&0&0&0&0&0&0&0&0\\
0&0&f_+&0&0&0&0&0&0&0&0&0&0&0&0&0\\
0&0&0&h&0&0&0&0&0&0&0&0&g&0&0&0\\
0&g&0&0&h&0&0&0&0&0&0&0&0&0&0&0\\
0&0&0&0&0&f_-&0&0&0&0&0&0&0&0&0&0\\
0&0&0&0&0&0&h&0&0&g&0&0&0&0&0&0\\
0&0&0&0&0&0&0&f_-&0&0&0&0&0&0&0&0\\
0&0&0&0&0&0&0&0&f_+&0&0&0&0&0&0&0\\
0&0&0&0&0&0&g&0&0&h&0&0&0&0&0&0\\
0&0&0&0&0&0&0&0&0&0&f_+&0&0&0&0&0\\
0&0&0&0&0&0&0&0&0&0&0&h&0&0&g&0\\
0&0&0&g&0&0&0&0&0&0&0&0&h&0&0&0\\
0&0&0&0&0&0&0&0&0&0&0&0&0&f_-&0&0\\
0&0&0&0&0&0&0&0&0&0&0&g&0&0&h&0\\
0&0&0&0&0&0&0&0&0&0&0&0&0&0&0&f_-
\end{array}\right).
\end{equation}
There are two solutions. The first one corresponds to
\begin{equation}
S^{(3,1)}=\{0,4,0,1,\theta,0,0\}.
\end{equation}
Here $f_+=f_-={\rm sinh}(\lambda+\eta)$ for $\theta=2\cosh\eta>2$,
$f_+=f_-={\rm sin}(\lambda+\eta)$ for $\theta=2\cos\eta<2$ and $f_+=f_-=\lambda+\eta$ for
$\theta=2$. The latter solution is related to the special case of (15) with $\alpha=\beta$ and
$\gamma=0$ as well to the special case of (17) with $\alpha=0$.

The second solution corresponds to
\begin{equation}
S^{(3,2)}=\{0,4,0,1,0,0,\theta\}.
\end{equation}
Here $f_{\pm}={\rm sinh}(\eta\pm\lambda)$ for $\theta=2\cosh\eta>2$,
$f_{\pm}={\rm sin}(\eta\pm\lambda)$ for $\theta=2\cos\eta<2$
and $f_{\pm}=\eta\pm\lambda$ for $\theta=2$.
In both the cases
$g={\rm sinh}\lambda$, $h={\rm sinh}\eta$ for $\theta>2$,
$g={\rm sin}\lambda$, $h={\rm sin}\eta$ for $\theta<2$,
and $g=\lambda$, $h=\eta$ for $\theta=2$.

\section{The group 4}
In this group $R$-matrices have the form,
\begin{equation}
R^{(4)}(\la)=\left(\begin{array}{cccccccccccccccc}
f_1&0&0&0&0&0&0&0&0&0&0&0&0&0&0&0\\
0&f_2&0&0&0&0&0&0&0&0&0&0&0&0&0&0\\
0&0&f_1&0&0&0&0&0&0&0&0&0&0&0&0&0\\
0&0&0&f_3&0&0&-g_1&0&0&g_2&0&0&g_1&0&0&0\\
0&0&0&0&f_2&0&0&0&0&0&0&0&0&0&0&0\\
0&0&0&0&0&f_1&0&0&0&0&0&0&0&0&0&0\\
0&0&0&-g_1&0&0&f_3&0&0&g_1&0&0&g_2&0&0&0\\
0&0&0&0&0&0&0&f_1&0&0&0&0&0&0&0&0\\
0&0&0&0&0&0&0&0&f_1&0&0&0&0&0&0&0\\
0&0&0&g_2&0&0&g_1&0&0&f_3&0&0&-g_1&0&0&0\\
0&0&0&0&0&0&0&0&0&0&f_1&0&0&0&0&0\\
0&0&0&0&0&0&0&0&0&0&0&f_2&0&0&0&0\\
0&0&0&g_1&0&0&g_2&0&0&-g_1&0&0&f_3&0&0&0\\
0&0&0&0&0&0&0&0&0&0&0&0&0&f_1&0&0\\
0&0&0&0&0&0&0&0&0&0&0&0&0&0&f_2&0\\
0&0&0&0&0&0&0&0&0&0&0&0&0&0&0&f_1
\end{array}\right),
\end{equation}
where $g_2=f_2-f_3$.

There are two solutions. The first one corresponds to
\begin{equation}
S^{(4,1)}=\{1,4\varepsilon,-4,-\varepsilon,1,0,0\},
\end{equation}
where $f_1=f_2=\sh(\la+\eta)$, $\varepsilon g_1=g_2=\sh\la$,
$f_3=\sinh(\lambda+\eta)-\sinh\lambda$, $\sh\eta=\sqrt{3}$. For $\varepsilon=1$ this model is the
special case of (15) with $\beta=\gamma=0$.
For $\varepsilon=-1$ it is $SU(2)\times SU(2)$-symmetric (in fact $SU(4)$-symmetric \cite{21}),
and as it was mentioned in \cite{11} corresponds to the four-critical point in the phase
diagram of the ferromagnetic $SU(2)\times U(1)$-symmetric spin-orbital model. This model was
also suggested in \cite{10} as a model of correlated electrons in a lattice of Berry phase
molecules. It was also shown in \cite{19} that it is a Temperly-Lieb system \cite{22}.

The second solution corresponds to,
\begin{equation}
S^{(4,2)}=\{2,4\varepsilon,-8,-\varepsilon,5,0,0\},
\end{equation}
where
$f_1=f_2{\rm e}^{\lambda}=4e^{2\la}-1,\quad f_3=2e^{\la}+e^{-\la},
\quad g_1=\varepsilon(1-e^{-2\la}),\quad g_2=4\sh\la$.

\end{document}